\documentclass{PoS}

\title{Global description of bottomonium suppression in proton-nucleus and nucleus-nucleus collisions at LHC energies}

\ShortTitle{Bottomonium suppression in $pA$ and $AA$ collisions at LHC energies}

\author{\speaker{Elena G. Ferreiro}\\
Departamento de F{\'\i}sica de Part{\'\i}culas and IGFAE, Universidade de Santiago de Compostela, 15782 Santiago de Compostela, Spain\\
Laboratoire Leprince-Ringuet, Ecole polytechnique, CNRS/IN2P3, Universit\'e Paris-Saclay, F-91128 Palaiseau, France\\
        E-mail: \email{elena@fpaxp1.usc.es}}

\abstract{We show that we can reach a global and coherent description
of bottomonium suppression in both proton-nucleus and nucleus-nucleus collisions at the LHC energies. 
The measured relative suppression of the excited bottomonium
states as compared to their ground state allows us to constrain the scattering cross
sections between the bottomonia and the comovers created during the collisions. Our result hints at a similar
energy distribution of these comovers in the environment created by proton-nucleus and nucleus-nucleus
collisions. Along the way of our study, we also update our knowledge for the bottomonium feed-down pattern
and our comprehension on the modification of the nuclear parton distribution functions in proton-nucleus collisions.
We present our results for $p$Pb collisions at 5.02 and 8.16 TeV and for PbPb collisions at 2.76 and 5.02 TeV.}

\FullConference{XXVI International Workshop on Deep-Inelastic Scattering and Related Subjects (DIS2018)\\
		16-20 April 2018\\
		Kobe, Japan}

\begin{document}

\section{Introduction}
The idea of quarkonium sequential suppression as a signature of the formation of a hot quark-gluon plasma (QGP) \cite{Matsui:1986dk} --different quarkonium states would melt sequentially at different temperatures within a deconfined medium--, was supported by the relative suppression of bottomonium observed by CMS Collaboration in lead-lead collisions~\cite{Chatrchyan:2011pe,Chatrchyan:2012lxa}: 
the excited $\Upsilon$(2S) and $\Upsilon$(3S) states were suffering more suppression than the ground $\Upsilon$(1S) state.

However, a couple of years later, the same Collaboration found a similar effect \cite{Chatrchyan:2013nza} --although of different magnitude-- in proton-lead collisions, where no QGP formation was expected. 
Moreover, a relative suppression of charmonium was equally found by PHENIX  \cite{Adare:2013ezl} and ALICE \cite{Abelev:2014zpa} Collaborations, also in asymmetric proton(deuteron)-nucleus collisions.
While at lower --SPS-- energies the relative suppression could be attributed to the interaction of the different quarkonium states --with different break-up cross sections-- with the remnants of the colliding nucleus, 
at higher energies the quarkonium formation time will be boosted. This implies that in this case the $Q\bar{Q}$ pair does not have the time to evolve into any physical state when it escapes the nuclear matter.
Thus, one cannot appeal to this mechanism to justify the observed difference, neither can one invoke initial-state effects --as the modification of the nuclear parton distribution functions (nPDFs)--  which should have a similar impact on the 
different states.

A natural explanation could be a final-state effect acting over sufficiently long time. According to the data, this effect should be stronger in the nucleus-going direction, and it should increase with centrality.
Already in the late 90's, an effect of this type, based on the interaction of the quarkonium with the comoving medium --{\it i.e.} made of particles with similar rapidities and whose density is directly connected
to the particle multiplicity-- was introduced. Here, we propose to explain the relative suppression of bottomonia in both $p$Pb and PbPb LHC collisions by this interaction.

\section{The improved comover interaction model (ICIM)}
Within this model \cite{Ferreiro:2012rq,Ferreiro:2014bia,Ferreiro:2018wbd},
quarkonia are suppressed due to scatterings
with the comoving particles.
The evolution equation that governs the density of bottomonium at a given transverse coordinate $s$, impact parameter $b$ and rapidity~$y$, $\rho^{\Upsilon}(b,s,y)$, follows the expression
\begin{equation}
\label{eq:comovrateeq}
\tau \frac{\mbox{d} \rho^{\Upsilon}}{\mbox{d} \tau} \, \left( b,s,y \right)
\;=\; -\langle\sigma^{co-\Upsilon}\rangle\; \rho^{co}(b,s,y)\; \rho^{\Upsilon}(b,s,y) \;,
\end{equation}
where $\langle \sigma^{\rm co -\Upsilon} \rangle$ corresponds to the interaction cross section of bottomonium with the comoving medium of transverse density~$\rho^{co}(b,s,y)$.
It depends essentially on 2 parameters, applicable to the entire bottomonium family. 
It reads
\begin{equation}
{\langle  \sigma^{\rm co-{\Upsilon}}  \rangle (T_{\rm eff},n) = 
\frac{\displaystyle\int_0^\infty dE^{\rm co }\,  {\cal P}(E^{\rm co };T_{\rm eff}) \, \sigma^{\rm co-{\Upsilon}}(E^{\rm co})}
{\displaystyle\int_0^\infty dE^{\rm co } \,  {\cal P}(E^{\rm co };T_{\rm eff})},  
{\rm where}\;
{\cal P}(E^{\rm co };T_{\rm eff}) \propto \frac{1}{e^{E^{\rm co }/T_{\rm eff}}-1}}
\end{equation}
corresponds to the Bose-Einstein energy distribution of the comovers, which introduces our first parameter, the effective temperature of these comovers $T_{\rm eff}$.

Our parametrisation of the energy dependence for the cross section $\sigma^{\rm co-{\Upsilon}}(E^{\rm co})$ interpolates from
$\sigma^{\rm co-{\Upsilon}}(E^{\rm co}=E^{\Upsilon}_{\rm thr})=0$ at threshold up to 
the geometrical cross section, 
$\sigma^{\rm co-{\Upsilon}}(E^{\rm co} \gg E^{\Upsilon}_{\rm thr})=\sigma^{\Upsilon}_{\rm geo}$,
where $\sigma^{\Upsilon}_{\rm geo} \simeq \pi r_{\Upsilon}^2$,
being $r_{\Upsilon}$ the bottomonium Bohr radius.
It reads 
\begin{equation}
\label{eq:sigma}\sigma^{\rm co-{\Upsilon}}(E^{\rm co}) =\sigma^{\Upsilon}_{\rm geo}  \left(1-\frac{E^{\Upsilon}_{\rm thr}}{E^{\rm co }}\right)^n
\end{equation}
where $E^{\Upsilon}_{\rm thr}=M_{\Upsilon}+m_{\rm co}-2 M_B$ is the threshold energy to break the bottomonium bound state and
$E^{\rm co}=\sqrt{p^2+m_{\rm co}^2}$ is the energy of the comover in the quarkonium rest frame. 
The second parameter of our modeling, $n$, characterises how quickly the cross section approaches
the geometrical cross section. We will let $n$ varying from 0.5 to 2.

In order to proceed with the fit, it is mandatory to take into account the feed-down contributions.
 The feed-down fractions \cite{Andronic:2015wma} for the $\Upsilon$(1S) can be estimated as: 70\% of direct $\Upsilon$(1S), 8\% from $\Upsilon$(2S) decay, 1\% from $\Upsilon$(3S),
15\% from $\chi_{\rm B1}$, 5\% from $\chi_{\rm B2}$ and 1\% from $\chi_{\rm B3}$, 
while for the $\Upsilon$(2S) the different contributions would be: 63\% direct $\Upsilon$(2S), 4\% of $\Upsilon$(3S), 30\% of $\chi_{\rm B2}$ and 3\% of  $\chi_{\rm B3}$.
For the  $\Upsilon$(3S), 40\% of the contribution will come from decays of  $\chi_{\rm B3}$.

\begin{figure}[thb]
\hspace{-0.5cm}
\includegraphics[width=0.9\linewidth]{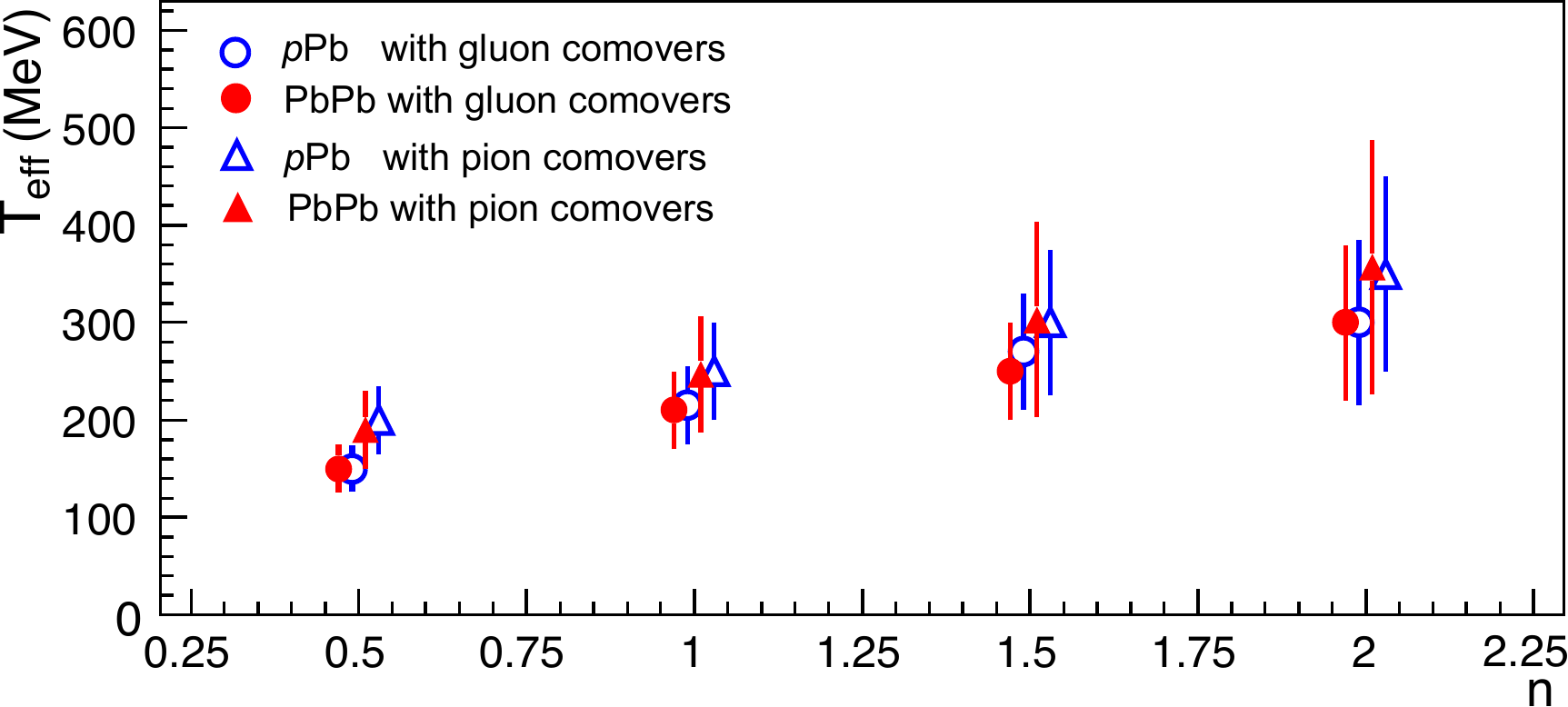}
\vskip -0.2cm
\caption{
\label{fig:figTvsn}
Fitted $T_{\rm eff}$ considering  pion (triangles) or gluon (circles) comovers from our fits to  $p$Pb (empty blue) and PbPb (filled red) data
for different $n$ from 0.5 to 2. [The points have been horizontally shifted for readability.]
}
\end{figure}

We have performed our fit of $T_{\rm eff}$ for different values of $n$ with both gluon ($m_{\rm co}=0$) or pion ($m_{\rm co}=0.140$ GeV) comovers. For $p$Pb collisions, we have used the CMS \cite{Chatrchyan:2013nza} and ATLAS \cite{Aaboud:2017cif} data at 5.02 TeV on relative nuclear suppression factors. For PbPb collisions, we have used the CMS data at 2.76 TeV~\cite{Chatrchyan:2012lxa} and at 5.02 TeV~\cite{Sirunyan:2017lzi}, also on relative nuclear suppression factors. Our results are shown in  Fig.~\ref{fig:figTvsn}.
All  the combinations yield to $T_{\rm eff}$s in the range 200 to 300 MeV for our assumed range for $n$. In what follows, our results will be shown for $n=1$ and $T_{\rm eff}=250 \pm 50$ MeV. Choosing different couples of $n$ and $T_{\rm eff}$ yield to very similar results since the variation of $n$ is compensated by that of $T_{\rm eff}$.

\section{Results}
Our results for the relative nuclear modification factors 
in $p$Pb collisions at 5.02 TeV are presented in Table~\ref{tab:pPb} 
together with the CMS \cite{Chatrchyan:2013nza} and ATLAS \cite{Aaboud:2017cif} experimental data. 
\begin{table}[hbt!]
\begin{center}\setlength{\arrayrulewidth}{1pt}
\begin{tabular}{cccc}
\hline\hline
& ICIM fit & Experimental values\\
\hline
 & $-1.93 < y < 1.93$ & CMS data\\
 $\Upsilon{\rm(2S)}/ \Upsilon{\rm(1S)}$ &0.91 $\pm$ 0.03 & 0.83 $\pm$ 0.05 (stat.) $\pm$ 0.05 (syst.)\\
  $\Upsilon{\rm(3S)}/ \Upsilon{\rm(1S)}$  & 0.72 $\pm$ 0.02  & 0.71 $\pm$ 0.08 (stat.) $\pm$ 0.09 (syst.)\\
 & $-2.0 < y < 1.5$ & ATLAS data\\
  $\Upsilon{\rm(2S)}/ \Upsilon{\rm(1S)}$ &0.90 $\pm$ 0.03 & 0.76 $\pm$ 0.07 (stat.) $\pm$ 0.05 (syst.)\\
  $\Upsilon{\rm(3S)}/\Upsilon{\rm(1S)}$  & 0.71 $\pm$ 0.02  & 0.64 $\pm$ 0.14 (stat.) $\pm$ 0.06 (syst.)\\
 \hline\hline
\end{tabular}
\caption{$R^{\Upsilon{\rm(nS)}/ \Upsilon{\rm(1S)}}_{pPb}$ at 5.02 TeV}\label{tab:pPb}
\end{center}
\end{table}

\begin{figure}[htb!]
\hspace{-0.2cm}
\centerline{\includegraphics[width=\linewidth]{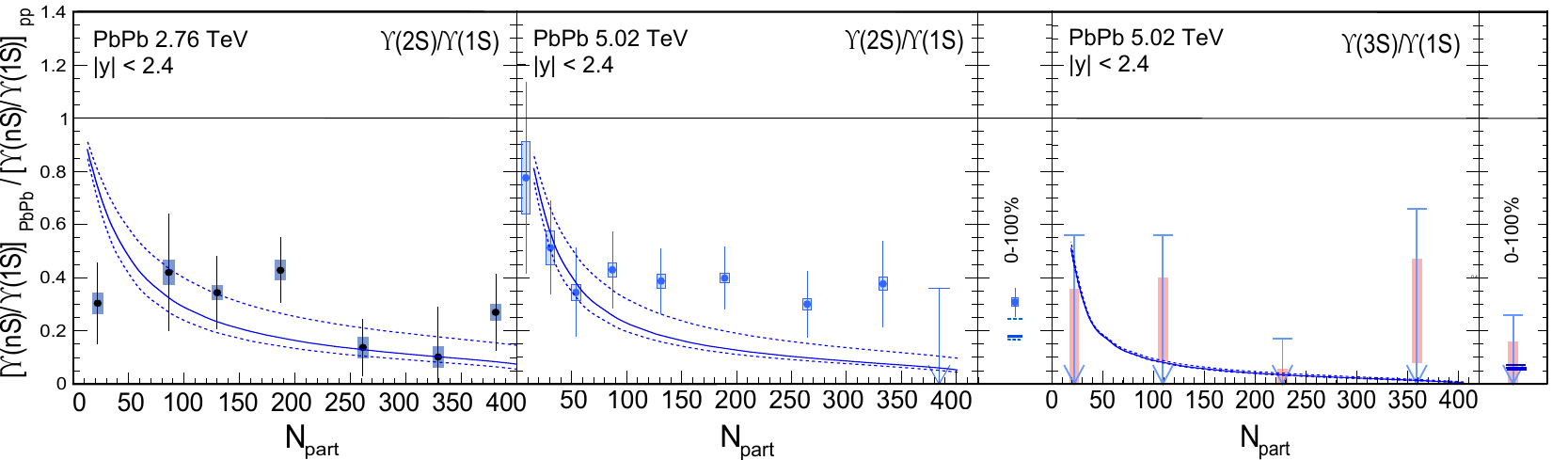}}
\caption{\label{fig:figdoubleratioPbPb}
The double ratio ($R^{\Upsilon{\rm(nS)}/ \Upsilon{\rm(1S)}}_{PbPb}$) for $\Upsilon$(2S) over $\Upsilon$(1S) at 2.76 and 5.02 TeV and $\Upsilon$(3S) over $\Upsilon$(1S) at 5.02 TeV
as a function of $N_{part}$ obtained from the ICIM and compared to the CMS data at 2.76 TeV \cite{Chatrchyan:2012lxa} and 5.02 TeV~\cite{Sirunyan:2017lzi}.
The dashed line depicts the uncertainty from the fit of $\sigma^{co-\Upsilon}$.
}
\end{figure}

Fig.~\ref{fig:figdoubleratioPbPb} shows our results on the centrality dependence of the relative suppression of $\Upsilon$(nS) at 2.76 and 5.02~TeV for PbPb collisions compared to the CMS data~\cite{Chatrchyan:2012lxa,Sirunyan:2017lzi}.

Once the parameters of our approach have been fixed with the relative suppression measurements, we can address
the absolute suppression of each state. In this case, 
other nuclear effects, which cancel in the double ratio of the excited-to-ground state suppression,
do not cancel anymore. At LHC energies, the main one is the nuclear modification 
of parton distribution functions. It is easily accounted for by using available global nPDF fits with uncertainties.

\begin{figure}[hbt!]
\centerline{\includegraphics[width=0.38\linewidth]{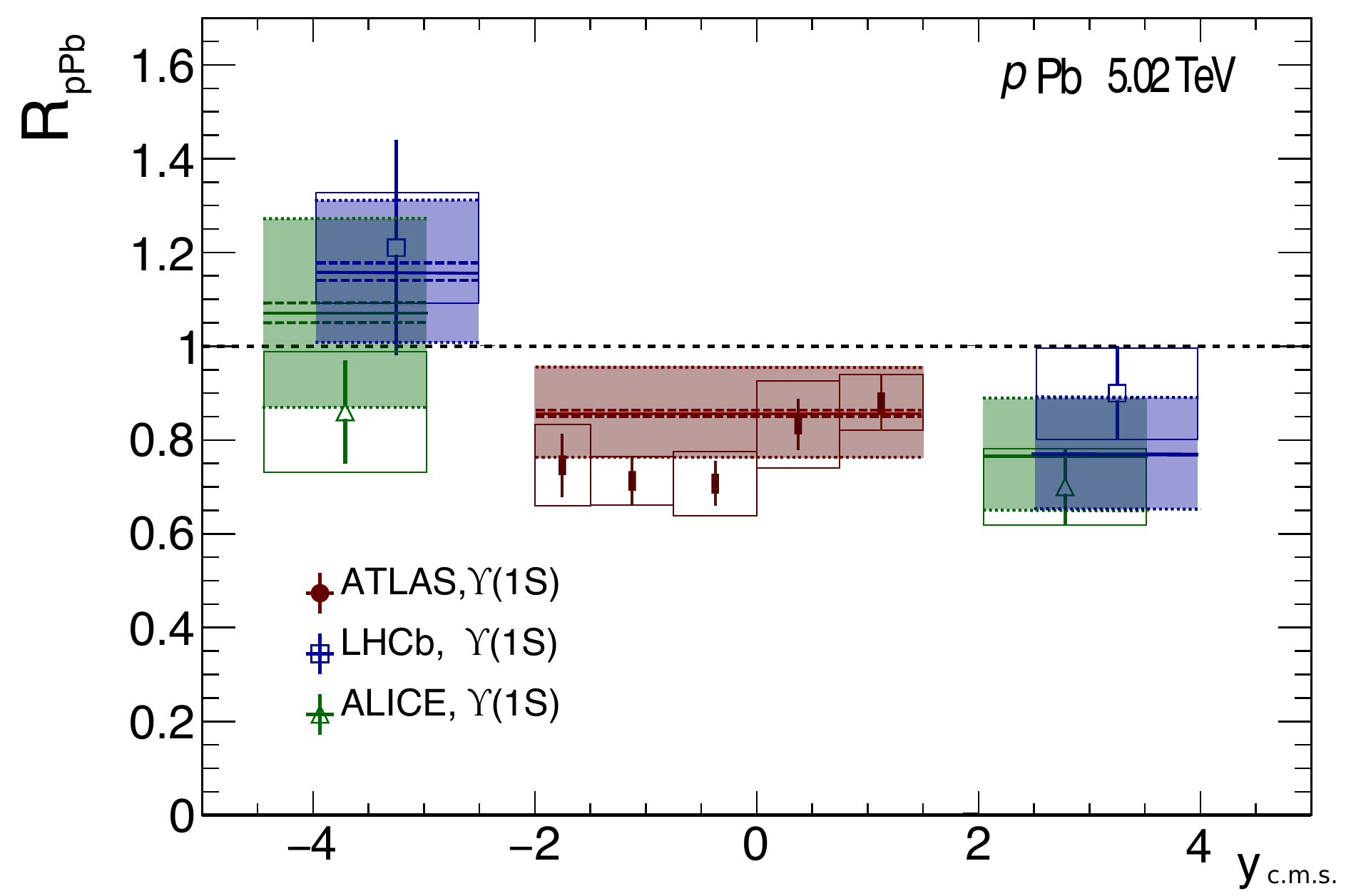}\includegraphics[width=0.29\linewidth]{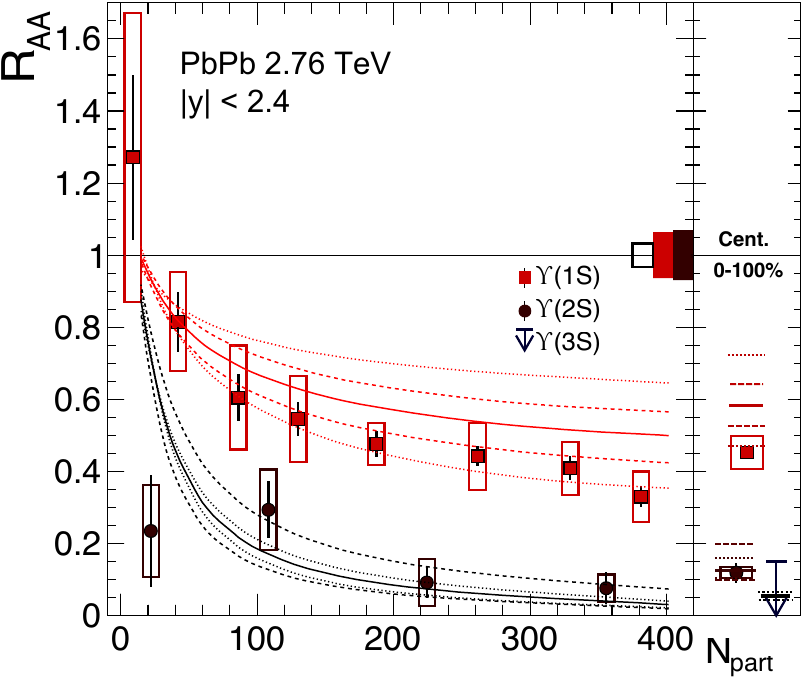}
\includegraphics[width=0.31\linewidth]{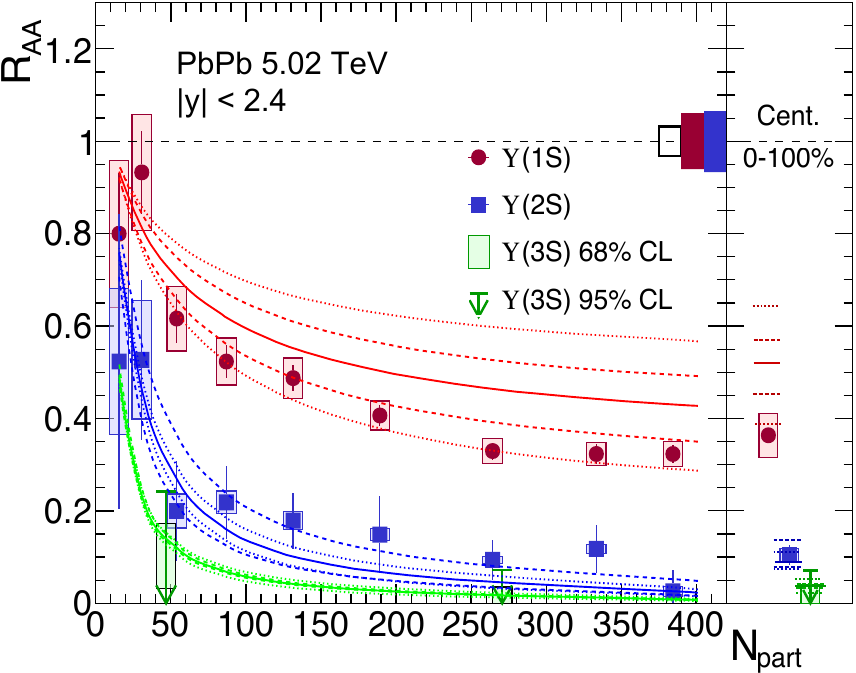}}
\caption{\label{fig:fig1SpPbandPbPb} 
$R_{p\rm Pb}^{\Upsilon(1S)}$ vs rapidity (right) at 5.02 TeV compared to the LHC data \cite{Aaij:2014mza,Abelev:2014oea,Aaboud:2017cif} and 
$R_{PbPb}^{\Upsilon(nS)}$ vs $N_{\rm part}$ (left) compared to the CMS data at 2.76 TeV \cite{Khachatryan:2016xxp} and 5.02 TeV~\cite{Sirunyan:2018nsz}.
The uncertainty from the fit of $\sigma^{co-\Upsilon}$ (dashed line) and from the nCTEQ15 shadowing (dotted line) are shown separately.
}
\end{figure}

In Fig.~\ref{fig:fig1SpPbandPbPb}, we
show $R_{p\rm Pb}^{\Upsilon(1S)}$ vs rapidity at $\sqrt{s}=5.02$~TeV compared to the available experimental 
data~\cite{Aaboud:2017cif,Aaij:2014mza,Abelev:2014oea} from ATLAS, LHCb and ALICE. An overall good agreement is obtained. 
Going to PbPb collisions, our results for the 3 $\Upsilon$ states at 2.76 and 5.02 TeV are also shown in Fig.~\ref{fig:fig1SpPbandPbPb} together with
the CMS data \cite{Khachatryan:2016xxp,Sirunyan:2018nsz}.
A good agreement is obtained in the 3 cases with the same parameters used to reproduce the relative suppression. The nCTEQ15 shadowing \cite{nCTEQ15} has been taken into account.

\begin{figure}[hbt!]
\hspace{0.3cm}
\centerline{\includegraphics[width=0.76\linewidth]{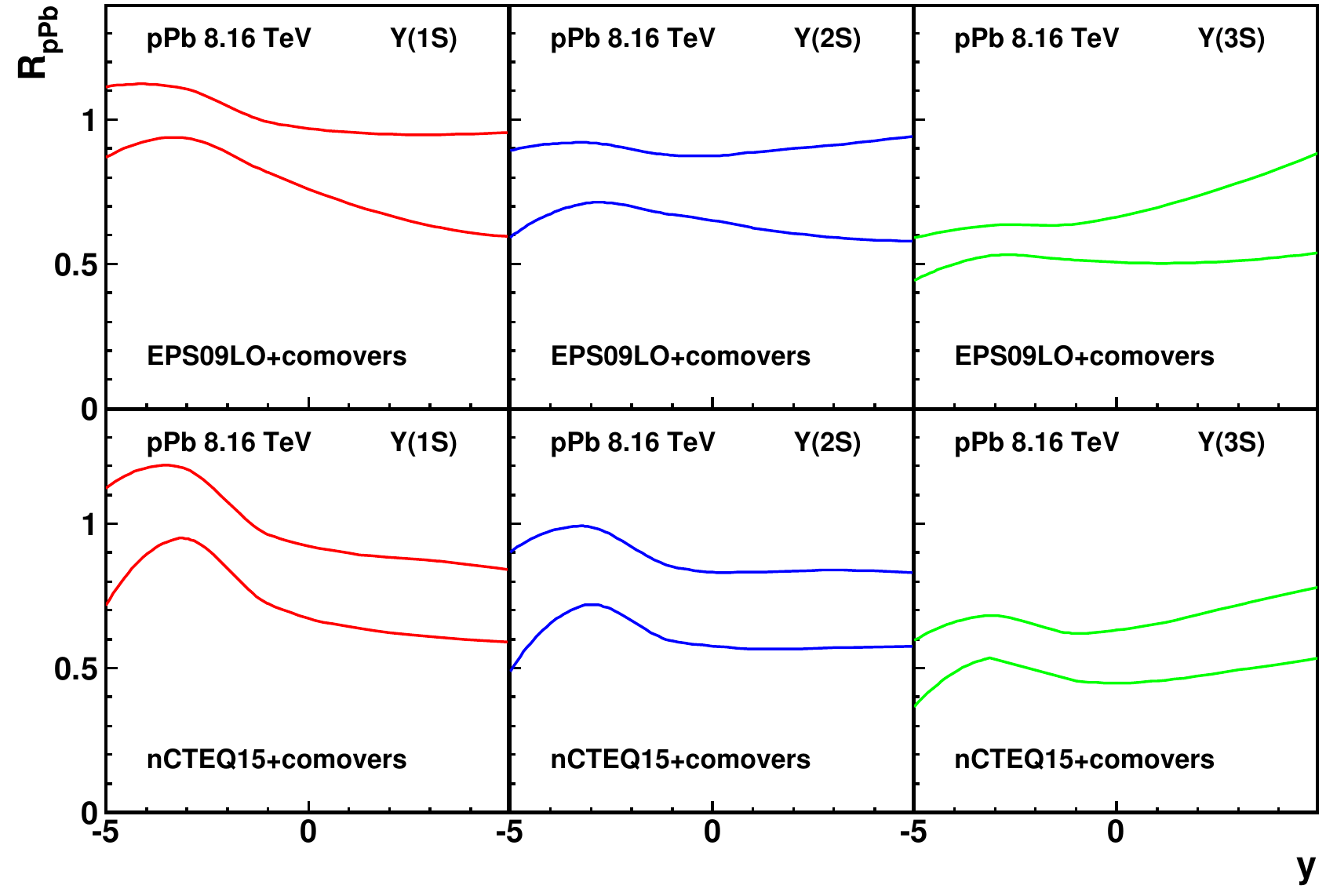}\includegraphics[width=0.36\linewidth]{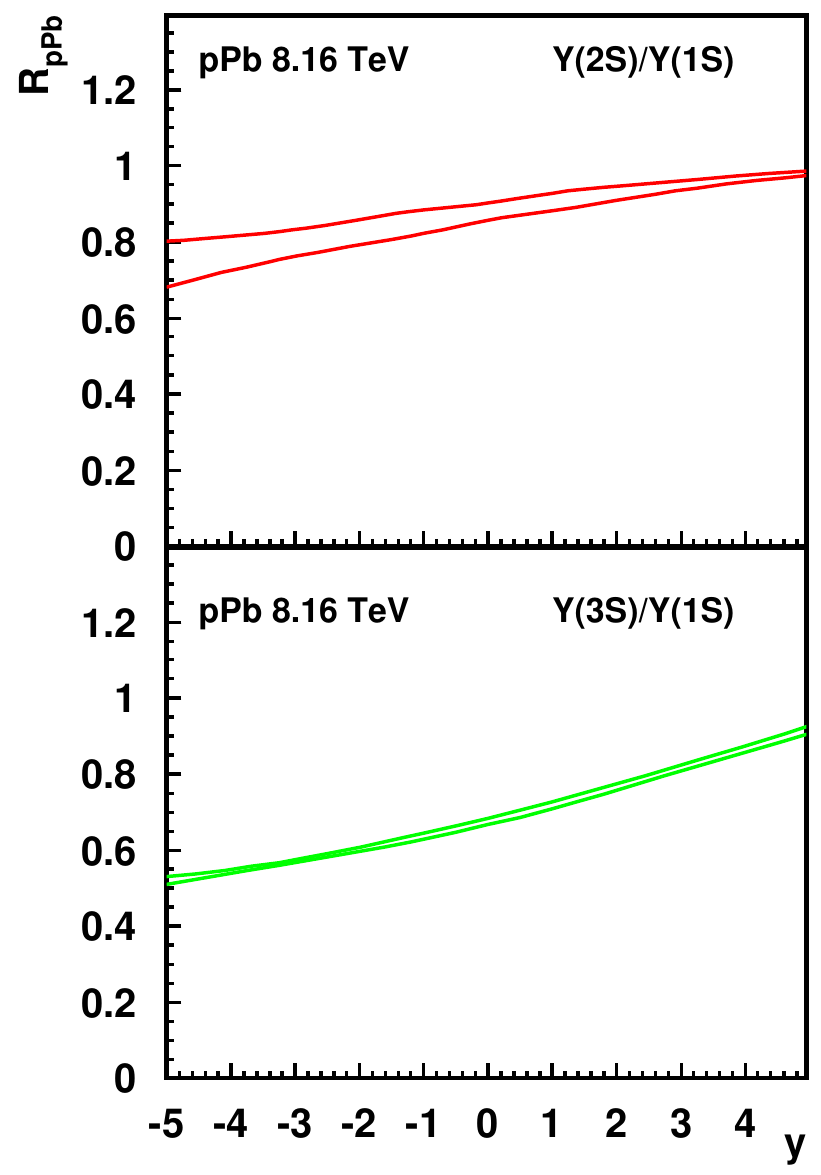}}
\caption{\label{fig:fignSpPb8TeV}
$R_{p\rm Pb}^{\Upsilon(nS)}$ vs rapidity at 8.16 TeV. Two different shadowing parametrisations (nCTEQ15 and EPS09LO) have been taken into account.
}
\end{figure}

To finish, in Fig.~\ref{fig:fignSpPb8TeV}, our predictions for $p$Pb collisions at 8.16 TeV are shown using two different parametrisations for the shadowing, nCTEQ15 \cite{nCTEQ15}  and EPS09LO \cite{Eskola:2009uj}.
Our results agree with the recent data obtained by LHCb Collaboration \cite{Aaij:2018scz}.

\section{Conclusions}
The relative suppression of the excited 
bottomonium states as compared to their ground state can be explained by the interaction with comovers.
We propose a generic formula for all the quarkonia states and suggest a connection with the 
energy distribution of the comovers, thus with an effective local temperature.
 This leads to a consistent magnitude for the 
$\Upsilon$ suppression
in both $p$Pb and  PbPb collisions
when combined with
shadowing.

The nature of comovers could correspond to 2 different scenarios.
If the comovers are identified with partons, the ICIM provides an effective modelling of bottomonium dissociation in the QGP.
If the comovers are identified with hadrons both in $pA$ and $AA$ collisions,  this would imply that the $\Upsilon$s remain unaffected by the presence of a
possible QGP.
Obviously, more complex --and intermediate-- scenarios could be proposed, where the suppression takes place
both with partons and hadrons. 


\end{document}